\documentclass[preprint,12pt]{elsarticle}

\usepackage{amsmath}
\usepackage{bm}
\usepackage{amsfonts} 
\usepackage{float}

\newcommand*\xbar[1]{%
  \hbox{%
    \vbox{%
      \hrule height 0.5pt 
      \kern0.5ex
      \hbox{%
        \kern-0.1em
        \ensuremath{#1}%
        \kern-0.1em
      }%
    }%
  }%
} 
\newtheorem{theorem}{Theorem}[section]
\newtheorem{lemma}[theorem]{Lemma}

\newtheorem{corollary}[theorem]{Corollary}

\newenvironment{proof}[1][Proof]{\begin{trivlist}
\item[\hskip \labelsep {\bfseries #1}]}{\end{trivlist}}
\newenvironment{definition}[1][Definition]{\begin{trivlist}
\item[\hskip \labelsep {\bfseries #1}]}{\end{trivlist}}

\usepackage{amssymb}

\usepackage[nodots]{numcompress}

\journal{arXiv}

\begin{document}


\begin{frontmatter}


\tnotetext[label1]{Corresponding author: Terence R. Smith, Department of Computer Science, University of California at Santa Barbara, Ca 93106}
\author{Terence R. Smith { }\fnref{label2}}
 \ead{smithtr@cs.ucsb.edu}

\title{
Natural Wave Numbers, Natural Wave Co-numbers,  
and the Computation of the Primes. }



\begin{abstract}

The paper exploits an isomorphism between the natural numbers 
$\mathbb{N}$
and a space $\mathbb{U}$ of periodic 
sequences of the roots of unity in constructing a recursive procedure
that represents the prime numbers.
The nth natural wave number ${\bm u}_n$ is the countable sequence 
of the nth roots of unity having frequencies k/n for all integer phases k.
The space $\mathbb{U}$ of natural wave numbers
is closed under a commutative and associative binary operation
${\bm u}_m\odot{\bm u}_n={\bm u}_{mn}$, termed the circular product,
and is isomorphic with $\mathbb{N}$
under their respective product operators.
Functions are defined on $\mathbb{U}$ 
that partition wave numbers into two complementary sequences, 
of which the co-number $ {\overset {\bm \ast }{ \bm u}}_n$ is a function of a wave number
in which zeros replace its positive roots of unity. 
It is shown, using circular products $ {\overset {\bm \ast }{ \bm U}}_{i}$ of the first i prime co-numbers,
that if ${p}_1,  ... , { p}_{i+1}$ 
are the first $i+1$ prime phases,
then the phases in the range $p_{i+1} \leq k < p^2_{i+1}$
that are associated with the non-zero terms of
$ {\overset {\bm \ast }{\bm U}}_{i}$
are, together with $ p_1, ...,p_i $, all of the prime phases 
less than $p^2_{i+1}$.
The recursive procedure 
$ {\overset {\bm \ast }{ \bm U}}_{N+1}=	 {\overset {\bm \ast }{ \bm U}}_{N}\odot{\overset {\bm \ast }{\bm u}}_{{N+1}}$
therefore represents prime numbers explicitly
in terms of preceding prime numbers, starting with the first
prime number $p_1=2$, and is shown never to terminate.
When applied with all of the primes identified
at the previous step, the recursive procedure identifies
approximately $7^{2(N-1)}/(2(N-1)ln7)$ primes 
at each iteration for $ N>1$.
When the phases of wave numbers are represented
in modular arithmetic, the
prime phases are representable
in terms of sums of reciprocals of the initial set of prime phases
and have a relation with the $\zeta$-function.
\end{abstract}



\end{frontmatter}


\section{Introduction}\label{sec1}

Elementary sieve methods for identifying prime numbers
(see, for example, Friedlander and Iwaniec (2010)\nocite{FI10})
may be characterized as procedures 
which, at every stage, select a periodic waveform propagating outward
from the origin of an axis labelled with the integers.
Beginning with a waveform of wavelenth two,
the wavelength of the next waveform selected is
equal to the integer of smallest magnitude
through which previous waveforms
have yet to propagate.
The wavelengths selected are the prime numbers.

The procedure implies that the identification of the primes
is an initial value problem that is consistent with their negative definition
as natural numbers that are not divisible by preceding prime numbers.
It leads to the problem of identifying the primes
as numbers that do not belong to countable sets
of multiples of the initial prime numbers.
Hence the initial value problem involves representing countable sequences
of numbers that have a wavelike structure and
numbers that are defined in terms of what they are not.
Although there are many solutions to this initial value problem
(see, for example, Montgomery and Vaughan (2012)\nocite{MV12}),
it has proven difficult to represent solutions
as functions of the initial prime values.

These difficulties suggest finding representations of the problem
in which wave-like sequences
of numbers are represented as numbers 
possessing an appropriate algebra
and in which there are numbers that are,
in some useful sense, complements of other numbers.
It is shown that such a representation may be 
based on defining a number system whose $n$th element
is a countable, periodic sequence of the $n$th
roots of unity. 
Such numbers, termed the natural wave numbers, 
are shown to possess a simple
algebra and to be capable of 
representing, in a usefully restricted sense, the complement of a wave number.
Products of these complements
are shown to be of use in identifying increasingly large
sets of prime numbers.
When the products are represented in terms of modular arithmetic,
one may derive representations, formed from sets of initial prime numbers,   
of following sets of prime numbers.

\section{The Natural Wave Numbers and their Operators}

The one-to-one correspondence 
between the natural numbers 
$\mathbb{N}$ and countable, periodic sequences of
the $n$th roots of unity
motivates 
\begin{definition}
A natural wave number 
of wavelength $n$ is a sequence 
\begin{eqnarray}
\label{eq:1}
{\bm u}_n =\left(  e^{2\pi i \frac{k}{n}  } 
{ \tiny \ \ \vert  \  k\ \epsilon\  \mathbb{Z}}, \ n\ \epsilon\  \mathbb{N}  \right),
\end{eqnarray}
in which ${\bm u}_n$ represents the countable, periodic sequence
of the nth roots of unity.
The number $k$,
corresponding to a term $e^{2\pi i \frac{k}{n}}$ of ${\bm u}_n$,
is termed a phase of
the wave number
and the number $\frac{k}{n}$ is termed a frequency.
\end{definition}

It is convenient to adopt the notation
\begin{eqnarray}
 _m{\bm u}_n 
=\left( e^{2\pi i \frac{k}{n}  } 
{ \tiny \ \  \vert  \ 1\leq  k \leq mn, \ n\epsilon \ \mathbb{N} }\right)
\end{eqnarray}
for the $nm$ elements of a sequence
for which $ 1 \leq k \leq mn$.
In particular, $_1{\bm u}_n $
is termed the principle part of the sequence ${\bm u}_n $
and $k=n$ is termed its principle phase.
The principal part of ${\bm u}_n $
may be visualized as
a unit circle centered on the origin of the complex plane,
with a circumference that
is marked, in a anticlockwise direction, with $n$ equi-spaced points 
corresponding to the first $n$ 
of the nth roots of unity with positive phases.
The space of wave numbers is denoted $\mathbb{U}$ and
the wave numbers may be linearly ordered 
in terms of $ n\  \epsilon \ \mathbb{N} $.

\subsection{Operators and Functions on the Space of Natural Wave Numbers}

Operators and functions defined on wave numbers
are assumed to act
in an element-wise manner 
on the exponential terms defining a wave number.

\subsubsection{The Unary Translation Operator }

A unary operator on ${\bm u}_n\  \epsilon\ \mathbb{U}$ circulates
the roots of unity around the circumference
of the unit circle. 
In particular, the operator 
\begin{eqnarray}
\label{eq:unary}
{\bm R}_j({\bm u}_n )
&=&  {\bm u}^j_n 
= \left(e^{2\pi i \frac{jk}{n} }\ 
\vert\   j,k\ \epsilon\  \mathbb{Z}   \right)  
\end{eqnarray}
translates the roots of unity
around the circumference
by raising them to a power $j \ \epsilon\  \mathbb{Z}$.

\subsubsection{The Circular Product Binary Operator}

The action of the circular product operator is specified in 
\begin{definition}
The application of the circular product operator 
$ { \odot}$ 
to wave numbers ${\bm u}_m$ and ${\bm u}_n$
results in a sequence whose principal part is
\begin{eqnarray}
\label{eq:cdp}
{_1}\left({\bm u}_m  \    \odot\ {\bm u}_n \right)\
= 
\left(_n{\bm u}_m \    \cdot _{m}{\bm u}_n
\right)^\frac{1}{{n+m}}
\end{eqnarray}
in which 
the operator
$\cdot$ represents the arithmetic, element-wise product operator
and the exponent $1/(n+m)$ 
represents an element-wise power
operator. 
\end{definition}
\noindent
In the case of ${\bm u}_2$ and ${\bm u}_3$, for example,
\begin{eqnarray}
{_1}\left({\bm u}_2 \    \odot {\bm u}_3\right)
&=&
\left(_3{\bm u}_2 \    \cdot {_2}{\bm u}_3
\right)^\frac{1}{2+3} 
=
{_1}{\bm u}_6.
\end{eqnarray}
and more generally one has 
\begin{theorem}
The space $\mathbb{U}$ of natural wave numbers
is closed under the direct circular product ${\odot}$.
\begin{proof}
\begin{eqnarray}
\left(_{n}{\bm u}_m \    \cdot {_{m}}{\bm u}_n
\right)^\frac{1}{{n+m}}
&=&
{_1}\left( (e^{2\pi i \frac{k}{m}  }\ \vert 
\ k\ \epsilon\ \mathbb{Z})\ { \cdot} \ 
(e^{2\pi i \frac{l}{n}  }\vert  \  
l\ \epsilon\ \mathbb{Z})\right)  ^\frac{1}{{n+m}}\nonumber  \\
&=&
{_1}( e^{2\pi i \frac{k}{m}  }\  { \cdot}\ 
e^{2\pi i \frac{k}{n}  }\ \vert  \  
k\ \epsilon\ \mathbb{Z})^\frac{1}{{n+m}} \nonumber   \\
&=&
{_1}(e^{2\pi i \frac{k({n+m})}{mn}  } \vert  \ 
\ k\ \epsilon\ \mathbb{Z})  ^\frac{1}{{n+m}} \nonumber \\
&=&
{_1}(e^{2\pi i \frac{k}{mn}  }\ \vert \ 
\ k=k\ \epsilon\ \mathbb{Z})  \nonumber \\
&=&{_1}{\bm u}_{mn}
\end{eqnarray}
and it follows that
${\bm u}_m  \    \odot\ {\bm u}_n \
=  {\bm u}_{mn}$.
\end{proof}
\end{theorem}
The circular product 
has an interpretation in terms of the rotations
of three unit circles whose centers
coincide with the origin
of the complex plane. The circles may be viewed as rotating 
in a clockwise direction at rates that
are inversely proportional, with the same constant
of proportionality, to the numbers $n, m, mn$
of their roots of unity.
They commence their rotation with each of their first roots of unity
($e^{2\pi i \frac{1}{m}},e^{2\pi i \frac{1}{n}},e^{2\pi i \frac{1}{mn}}$)
aligned at a point $p$ on the circumference of a fourth coincident unit circle
that does not rotate. As a result of their speeds of rotation,
each of their successive roots of unity coincide at the point $p$, 
because the time to rotate from one
root to the next is the same for all three circles.

The circular product is
commutative and associative as a result of its element-wise
definition and the commutativity and associativity of the arithmetic addition and multiplication operators. 
One may therefore generalize
its definition to products of $N$ wave numbers
\begin{eqnarray}
\label{eq:CP0}
	\underset {i=1} {\overset{ N} {   \odot} }{\bm u}_{n_i}
&=& 	 
\left({_{P/n_1}}{\bm u}_{n_1} \cdot \   _{P/n_2}{\bm u}_{n_2} \ \cdot \  ... \ 
 \cdot \ {_{P/n_N}}{\bm u}(n_N)
\right)^\frac{1}{{\sum_i(P/n_i)}}  
=\ {_1}{\bm u}_P
\end{eqnarray}
in which
$P=\overset {N} {\underset {1=1}\Pi}  {n_i}$.

It is clear that an invertible mapping
$M:\mathbb{N}\rightarrow\mathbb{U}$ exists between the natural numbers
and the natural wave numbers. 
Furthermore, one has
\begin{theorem}
The space $\mathbb{N}$ of natural numbers and the space
$\mathbb{U}$ of natural wave numbers
are isomorphic under their product operators $\cdot$ and $\odot$.
\begin{proof}
Let $M:\mathbb{N}\rightarrow\mathbb{U}$
be the invertible mapping  from $\mathbb{N}$ onto $\mathbb{U}$, 
then $M(m)\odot M(n)={\bm u}_{n}\odot{\bm u}_{m}
={\bm u}_{n\cdot m}=M(n\cdot m)$.
\end{proof}
\end{theorem}

\subsection{Functions of a Wave Number }

It is useful to specify functions of a wave number in terms of
\begin{definition}
A function of a wave number
${\bm u}_n\  \epsilon\  \mathbb{U}$ is
a periodic sequence 
\begin{eqnarray}
{\bm f}({{\bm u}_n )} 
=\left( { f}\big(  e^{2\pi i \frac{k}{n}  }\big )\  \vert
\  k\ \epsilon\  \mathbb{Z}, \  n\ \epsilon\  \mathbb{N}   \right) 
\end{eqnarray}
 defined 
by the  element-wise application of a function $f: \mathbb{C} \rightarrow  \mathbb{C}$.
\end{definition}
A function may be viewed as 
a one-to-one mapping between the phase $k$ of
the wave number and the value ${ f}\big(  e^{2\pi i \frac{k}{n}  }\big )$.
Since the arithmetic product and addition operators, $\cdot$ and $+$,
are applicable to pairs of function values,
the circular product operator is applicable to functions 
of wave numbers.

\subsubsection{
Decomposition Functions and the Co-numbers}

It is useful to define functions of a natural wave number ${\bf u}_n$
that, together, form a partition of the number.
Of particular value are the functions $ \overset{\bm \circ}
{\bm u}_n$ and $ \overset{\bm \ast}{\bm u}_n$ 
having
\begin{definition}
The principal parts of the $^\star$ and $^\circ$ functions are
\begin{eqnarray} 
\label{eq:fdef}
 {_1}\overset{\bm \ast}{\bm u}_n&=&
\left(  e^{2\pi i \frac{1}{n}  },
\ e^{2\pi i \frac{2}{n}  },..., \ e^{2\pi i \frac{(n-1)}{n}  }  ,\ 0 \right) 
\nonumber \\
{_1}{\overset{\bm \circ}{\bm u}}_n&=&
\big(\  0 ,\  0,..., 0 ,\ e^{2 i\pi \frac{n}{n}} \big) = 
\big(\  0 ,\  0,..., 0 ,\ 1 \big),
\end{eqnarray} 
\end{definition}
\noindent
which may be viewed, respectively, as mapping the positive real elements 
of ${\bf u}_n$ into zero and the complex and negative real values
of ${\bf u}_n$ into zero.

The sum of the functions
is
\begin{eqnarray}
\label{part}
\overset{\bm \circ}{\bm u }_n+\overset{\bm \ast}{\bm u }_n
={\bf u}_n
\end{eqnarray} 
in which $+$ is the arithmetic addition operator applied in
an element-wise manner.
Since the principal part 
$_1{\overset {\bm \circ}{\bm {u}}_n}=(0,...,0,1)$ may be viewed 
as a minimal representation of ${\bm u}_n$,
and since
$
{ \overset{\bm \ast}{\bm {u}}_n}  \odot  {\overset{\bm \circ}{\bm {u}}_n}
={\bf 0} 
$,
it is reasonable to view $ \overset{\bm \ast}{\bm {u}}_n$
as a complement of ${\bm {u}}_n$
and to make
\begin{definition}
The function $ \overset{\bm \ast}{\bm {u}}_n
=\left( 
(1-\delta_{k,jn})  e^{2\pi i \frac{k}{n}  } 
{ \tiny \ \ \vert \ j\ \epsilon\  \mathbb{Z}, 
k\ \epsilon\  \mathbb{Z}}, \ n\ \epsilon\  \mathbb{N}  \right)$,
in which $\delta_{k,jn}$ is the Kronecker delta,
is termed the co-number of ${\bm {u}}_n$.
\end{definition}
The partition (\ref{part}) leads to partitions of 
circular products of wave numbers
\begin{eqnarray}
 {_1}\left({\bm u}_m  \    \odot\ {\bm u}_n \right)
&=& 
\left({_{n}}\left( { {\overset{\bm \circ} { \bm u}_m} } \odot { {\overset{\bm \ast} 
{ \bm u}_m}}\right)\cdot
 {_{m}}{  \left( { {\overset{\bm \circ} { \bm u}_n} } \odot { {\overset{\bm \ast} 
{ \bm u}_n}}\right)}\right)^\frac{1}{n+m} \nonumber  \\
&=& _1\left( { {\overset{\bm \circ} { \bm u}_m} } \odot { {\overset{\bm \circ} 
{ \bm u}_n}}\right) 
+\  {_1}\left( { {\overset{\bm \circ} { \bm u}_m} } \odot { {\overset{\bm \ast} 
{ \bm u}_n}}\right) \nonumber \\
&+&\ _1\left( { {\overset{\bm \ast} { \bm u}_m} } \odot { {\overset{\bm \circ} 
{ \bm u}_n}}\right) 
+\  {_1}\left( { {\overset{\bm \ast} { \bm u}_m} } \odot { {\overset{\bm \ast} 
{ \bm u}_n}}\right) 
\end{eqnarray}
in which the term
 ${_1}\left( { {\overset{\bm \ast} { \bm u}_m} } \odot { {\overset{\bm \ast} 
{ \bm u}_n}}\right) $ is of particular interest since it involves all phases of
${ \bm u}_m$ 
and ${ \bm u}_n$ 
except $k=m$ and $k=n$ as, for example, in the case
\begin{eqnarray}
 {_1}\left( { {\overset{\bm \ast} { \bm u}_2} } 
\odot { {\overset{\bm \ast} 
{ \bm u}_3}}\right) 
=(e^{2\pi i \frac{1}{6}},0,0,0,e^{2\pi i \frac{5}{6}},0)
\end{eqnarray}
which, one notes, is not equal to $ {_1}\overset{\bm \ast} { \bm u}_6
=(e^{2\pi i \frac{1}{6}},e^{2\pi i \frac{2}{6}},e^{2\pi i \frac{3}{6}},
e^{2\pi i \frac{4}{6}}, e^{2\pi i \frac{5}{6}},0)$. 

\section{The Prime Natural Wave Numbers}

One may formulate the primality
of natural wave numbers
in terms of
\begin{definition} 
A natural wave number ${\bm u}_{n}$ is prime iff the number $n$
representing its wavelength and
principal phase is a prime number.
\end{definition}
\noindent
Hence there is a one-to-one correspondence between the prime wave numbers
and the prime numbers.
An analogue of the Fundamental Theorem of Arithmetic
for natural wave numbers is stated in
\begin{theorem}
If the natural number $n=p_1^{\alpha_1}...p_N^{\alpha_N}$ for some $N$,
then the natural wave number 
\begin{eqnarray}
{\bm u}_{n}=
{\bm u}_{p_1^{\alpha_1}...p_N^{\alpha_N}}=
{ \overset{N} {\underset {i=1} \odot}}
{\bm {u}}_
{p^{\alpha_i}_i }= { \overset{N} {\underset {i=1} \odot}}
\left({ \overset{\alpha_i} {\underset {j=1} \odot}}
{\bm {u}}_{p_i }\right)
\end{eqnarray}
\begin{proof}
By the definition of the circular product. 
\end{proof}
\end{theorem}
.

\section{Co-Numbers and the Recusive Identification and Representation of the Primes }

The cumulative circular products 
of the ${N-1}$ natural wave co-numbers 
${\overset {\bm \ast }{\bm u}_2,
...,\overset {\bm \ast }{\bm u}}_{N}$
are of value in identifying the prime numbers.
The circular product
\begin{eqnarray}
\label{eq:CP0}
{\overset {\bm \ast }{\bm U}}_N\big(k\big)
=
	\Big(\underset {n=2} {\overset{ N} {   \odot} } {\overset {\bm \ast }{\bm u}_n}\Big)
\big(k\big)
\end{eqnarray}
is a function of the phases $k$ of the
circular product 
${\bm U }_{N}\big(k\big)
=\underset {n=2} {\overset{ N} {   \odot} } {\bm u}_n
\big(k\big)$
of the first ${N-1}$ wave numbers.
The principal part of the domain of the function
may be visualized as the unit circle in the complex plane
defined by the $P_N=\underset {n=2} {\overset{ N} {   \Pi}} n_i$ roots of unity, with
each phase $k$ corresponding 
to an exponential term $e^{2\pi i k/P_{N}}$ of
the circular product ${\bm U }_{N}\big(k\big)$.

It follows immediately from the definitions of the
natural wave co-numbers ${\overset {\bm \ast }{\bm u}_2
...,\overset {\bm \ast }{\bm u}_{N}}$,
and of their circular product (\ref{eq:CP0}), 
that one has 
\begin{theorem}
\label{L1}
The function ${\overset {\bm \ast }{\bm U}}_N\big(k\big)
=\Big(\underset {n=2} {\overset{ N} {   \odot} } {\overset {\bm \ast }{\bm u}_n}\Big)
\big(k\big)
=0$
iff the phase $k$ is a multiple of any of the phases $k=2,...,N$. 
\begin{proof}
The value of the function for any phase $k$ is 
an arithmetic product of exponential terms
and zeros. Such a product is zero iff it contains at least one zero
which occurs iff $k$ is a multiple of one of the first $N$ phases.
\end{proof}
\end{theorem}
\noindent
The function values are therefore non-zero for any phase
$k$ iff it is prime relative to the phases $k=2,...,N$.
In the case of $N=3$, for example, the domain of the function
is ${\bm U}_{3}\big(k\big)={\bm u}_{6}(k)$ and the function values
for the phases of the first six wavelengths of the product are
\begin{eqnarray}
\label{eq:2,3,6}
{\overset {\bm \ast }{\bm U}}_3\big(k\big)
&=&
{_6}\big({\overset {\bm \ast }{\bm u}_2} \odot {\overset {\bm \ast }{\bm u}_3}\big)\big(k\big)  
= {_6}\big(\big( {  {\overset {\bm \ast }{{_3}{\bm u}}}_2} \cdot 
 {   {  {\overset {\bm \ast }{{_2}{\bm u}}}}_3}\big)^\frac{1}{5}\big) \big(k\big)\\
&=&\   \big(e^{2\pi i\frac{1}{6}}, 0,0, 0, e^{2\pi i\frac{5}{6}}, 0,
e^{2\pi i\frac{7}{6}}, 0,0, 0, e^{2\pi i\frac{11}{6}}, 0, \nonumber\\
&\ &\  \  e^{2\pi i\frac{13}{6}}, 0,0, 0, e^{2\pi i\frac{17}{6}}, 0,
e^{2\pi i\frac{19}{6}}, 0,0, 0, e^{2\pi i\frac{23}{6}}, 0,\nonumber\\
&\ &\ \ e^{2\pi i\frac{25}{6}}, 0,0, 0, e^{2\pi i\frac{31}{6}}, 0,
e^{2\pi i\frac{37}{6}}, 0,0, 0, e^{2\pi i\frac{43}{6}}, 0
\big)  \nonumber
\end{eqnarray}
in which the first non-prime phase is $5^2=25$
which, one notes, is the square of the first prime number following $2$ and $3$

\subsection{The Circular Product of Prime Co-numbers}

In identifying the phases associated with the zeros and non-zeros of 
${\overset {\bm \ast }{\bm U}}_N\big(k\big)$,
there is
no loss of generality in using circular 
products of the prime
wave co-numbers instead of the natural wave co-numbers
as long as they are defined over consistent ranges.
Denoting by $\lfloor N \rfloor_p$ the largest prime number
less than or equal to $N$ and by $\pi_p=\pi(\lfloor N \rfloor_p)$
the value of the prime counting function, 
the circular product of prime wave co-numbers
\begin{eqnarray}
\label{eq:CP0b}
\overset {\bm \ast }{\bm U}_{\pi_p}\big(k\big)=
	\underset {i=1} {\overset{ \pi_p} {   \odot} } {\overset {\bm \ast }{\bm u}}_{p_i}
\big(k\big), 
\end{eqnarray}
is equivalent to 
${\overset {\bm \ast }{\bm U}}_{N}\big(k\big)=
	\underset {n=2} {\overset{ N} {   \odot} } {\overset {\bm \ast }{\bm u}}_n
\big(k\big)$ in terms of the phases of its zero values 
\begin{lemma}
The phases associated with the zeros of the circular product of prime wave co-numbers
$\underset {i=1} {\overset{ \pi_p} {   \odot} } {\overset {\bm \ast }{\bm u}}_{p_i}(k)$
and the phases associated with the zeros of the product $
\underset {n=2} {\overset{ N} {   \odot} } 
{\overset {\bm \ast }{\bm u}}_{n}(k)$ of natural wave co-numbers
are identical.
\end{lemma}
\begin{proof}
By the definition of the circular product of natural wave co-numbers
$	\underset {n=2} {\overset{ N} {   \odot} } 
{\overset {\bm \ast }{\bm u}}_{n}(k)$,
the co-number for any multiple of a prime phase is preceded by the co-number for the prime phase.
Furthermore the phases of the zeros 
of a co-number whose principal phase is a multiple
of a preceding prime phase
form a subset of the phases of the zeros of the preceding prime co-number
as, for example, in the case 
${_2}{ \overset {\bm \ast }{\bm u}}_2=
(e^{2\pi i\frac{1}{2}}, 0, e^{2\pi i\frac{3}{2}} ,0)$
which precedes
$ {_1}{\overset {\bm \ast }{\bm u}}_4=
(e^{2\pi i\frac{1}{4}}, e^{2\pi i\frac{2}{4}}, e^{2\pi i\frac{3}{4}} ,0)$.
Hence the occurrence of any non-prime co-number in
$\underset {n=2} {\overset{N} {   \odot} } 
{\overset {\bm \ast }{\bm u}}_{n}(k)$ 
following the occurrence of
a prime factor co-number has no effect on which
phases are associated with zeros.
\end{proof}

It follows that one may employ
circular products of prime numbers
in identifying both the prime wave numbers and the prime natural numbers.
Before stating and proving a theorem concerning
their identification for general values of $N$,
it is helpful to examine special cases
of $N$.
Employing the notation
$	{\overset {\bm \ast }{\bm U}}_{P_N}(k)=
\underset {i=1} {\overset{ N} {   \odot} } {\overset {\bm \ast }{\bm u}}_{p_i}(k)$
for circular products of prime co-numbers,
one obtains for the first six wavelengths of the product for $N=1$
\begin{eqnarray}
{_6}{\overset {\bm \ast }{\bm U}}_1(k)&=&
{_6}\overset {\bm \ast}  {\bm {u}}_{2}(k) \\
&=& \big(e^{2\pi i\frac{1}{2}},  0, e^{2\pi i\frac{3}{2}}, 0,
e^{2\pi i\frac{5}{2}}, 0, e^{2\pi i\frac{7}{2}}, 0,
e^{2\pi i\frac{9}{2}}, 0, e^{2\pi i\frac{11}{6}},0\big),
\nonumber
\end{eqnarray}
whose function values may be viewed as
alternately occurring at the two roots of unity $(-1,0)$ and $(1,0)$
on the product circle of ${\bm u}_2$
in the complex plane.
The values
include all prime phases $3 \leq k < 3^2$,
with the first non-prime phase occurring at $k=3^2$.

The function values for the case $N=2$,
corresponding to
${\overset {\bm \ast }{\bm U}}_2(k) =
{\overset  {\bm \ast}{\bm u}_2} \odot {\overset  {\bm \ast}{\bm u}_3}$,
are displayed in equation (\ref{eq:2,3,6}). They
include all prime phases $5 \leq k < 5^2$,
with the first non-prime phase occurring at $k=5^2$.
It may be noted, however, that the identification of three prime phases
in addition to the prime phase $2$ for the case $N=1$
suggests that one might have proceeded directly to the case $N=3$ and 
to the circular
product ${\overset {\bm \ast }{\bm U}}_3(k)=
{\overset  {\bm \ast}{\bm u}_2} \odot {\overset  {\bm \ast}{\bm u}_3}
\odot {\overset  {\bm \ast}{\bm u}_5}$
while using the knowledge that the square of the 
next prime phase is
$ {7^2= 49}$ to
limit the sequence to phases that are prime.
It is straightforward to show that this leads to the identification
of the prime phases $7 \leq k < 7^2$,
with $k=49$ being the first non-prime phase.

\subsection{
Recursive Identification of Increasingly-Large Sets of Primes}

The preceding special cases suggest 
\begin{theorem}
If $ p_1,  ..., p_{N+1}$ are the first $N+1$ prime phases,
then the phases in the range $p_{N+1} \leq k < p^2_{N+1}$
that are associated with the non-zero terms of the circular product
of the first $N$ prime co-numbers
\begin{eqnarray}
 \overset {\bm \ast }{\bm {\bm U}}_{P_N}
= \underset {i=1} {\overset{ {N}}  {   \odot} }
 \overset {\bm \ast } {\bm {u}_{p_i} },
\end{eqnarray}
are, together with $ p_1,  ..., p_{N}  $, 
all of the prime phases 
less than $p^2_{N+1}$.

\begin{proof}
Let 
$ {\bm {U }}_{P_N}(k)$ be the circular product
of the first $N$ prime wave numbers
and let
 $
 \overset {\bm \ast }{\bm {\bm U}}_{P_N}(k)
= \underset {j=1} {\overset{ N}  {   \odot} }
 \overset {\bm \ast } {\bm {u}}_{p_j} (k)$ 
be function values defined on the phases of $ {\bm U}_{P_N}(k)$.
By Lemma \ref{L1},
$\overset {\bm \ast }{\bm {\bm U}}_{P_N}(k)=0$ 
iff the phase $k$ is a multiple of any of the prime numbers 
$p_1,...,p_N$.
The phase $k=p^2_{N+1}$ is not prime and is associated with
a non-zero value since it does not satisfy
this condition.
Any phase smaller than $k=p^2_{N+1}$ may be represented
by the Fundamental Theorem of Arithmetic
\begin{eqnarray} 
 k= p_1^{a_1}\cdot p_2^{a_2} \cdot\cdot\cdot p_N^{a_N} 
\cdot {p_{N+1}}^{a_{N+1}}\cdot\cdot\cdot 
 \left(\lfloor p^2_{N+1}\rfloor_p\right)^{a_m}
\end{eqnarray}
for some $m$,
in which $ \lfloor n\rfloor_p$
represents the largest prime phase that is less than $n$.
Since the function values of phases that contain 
any prime factors less than $p_{N+1}$
are zero
and since permissible phases are less than $p^2_{N+1}$, it follows 
that phases associated with non-zero function values
must have representations
\begin{eqnarray} 
 k= p^{s_1}_{N+1}\cdot\cdot\cdot 
 \lfloor p^2_{N+1}\rfloor^{s_m}_p
\end{eqnarray}
for some $m$ in which at most one of the $s_i$'s can take on the value one.
It follows that any non-zero phase that is less than $p_{m+1}^2$
must be a prime number in the range 
${p}_{N+1},...,\lfloor p^2_{N+1}\rfloor_p$. 
\end{proof}
\end{theorem}
Theorem $5.3$ implies a recursive procedure
for determining the primes, since
\begin{eqnarray}
   \overset {\bm \ast }{\bm {\bm U}}_{P_{N+1}}(k)
= \overset {\bm \ast }{\bm {\bm  U}}_{P_N}(k)  
\  {   \odot} \ 
 \overset {\bm \ast } {\bm {u}}_{p_{N+1}}(k)  ,
\end{eqnarray}
which is defined entirely in terms of the initial condition
$   \overset {\bm \ast }{\bm {\bm  U}}_{1}(k)
= \overset {\bm \ast } {\bm {u}}_{2}$.
Furthermore, this procedure leads to
the identification of every prime by
\begin{corollary}
The recursive procedure that is defined by
applying Theorem $5.3$ sequentially for $N=1,2,3,...$
does not terminate.
\begin{proof}
Theorem $5.3$ implies that in order to identify prime phases
at step $N+1$, it is necessary at step $N$ to  identify $p_{N+1}$
in order to construct $ {\overset {\bm \ast }{\bm  U}}_{N_{P+1}}$ 
and to identify $p_{N+2}$ in order to ensure the primality
of the numbers generated by $ {\overset {\bm \ast }{\bm  U}}_{N_{P+1}}$.
As shown above, this is the case for the first three steps $N\leq3$.
Assume it to be true for the $N$th step and that $p_{N+1}$ 
and $p_{N+2}$ are known.
Theorem $5.3$ also implies that all prime phases $k< p_{N+2}^2$
are identified by 
$ {\overset {\bm \ast }{\bm U}}_{N_{P+1}}$,
and hence $p_{N+2}$ is identified
at the $(N+1)$st iteration.
By Bertrand's Theorem, $p_{N+3} \leq 2 p_{N+2}$,
and since $2 p_{N+2} \leq  p^2_{N+2}$ for $p_{N+2}>2$,
$p_{N+3}$ is determined.
\end{proof}
\end{corollary}

Theorem ${ 5.3}$ implies that one may identify
increasingly-large sets of maximum size of prime numbers
in an iterative process by which, at the $N$th step,
one employs the maximum number
of correct primes
determined in the previous steps to compute
the following largest, error-free set of primes.
In particular, one has
\begin{corollary}
The largest prime phase
in the set of prime phases that may be correctly identified at each application of
Theorem ${ 5.3}$ is 
\begin{eqnarray}
\lfloor 7 \rfloor_p, \ \ 
\lfloor 7^2 \rfloor_p,\ \ 
\lfloor \lfloor 7^2 \rfloor_p^2 \rfloor_p ,\ \
\lfloor \lfloor \lfloor7^2 \rfloor^2_p \rfloor^2_p \rfloor_p,\ \  
\lfloor\lfloor\lfloor \lfloor7^2 \rfloor^2_p \rfloor^2_p\rfloor^2_p \rfloor_p ,\ \ 
\lfloor\lfloor \lfloor\lfloor7^2 \rfloor^2_p \rfloor^2_p\rfloor^2_p \rfloor^2_p\ ,...
\end{eqnarray}
\begin{proof}
In the first step at $N=1$, 
${\overset {\bm \ast } {\bm U }}_1
={\overset {\bm \ast }{\bm u }}_2$ and
it follows from equation (21) that the largest prime phase 
found is $7$. From Theorem {5.3} it follows that
the largest correctly identifiable prime phase
at each iteration for $N\geq 2$ is the largest prime phase
that is less than the square of the next known prime phase.
Hence for $N=2$, the next prime phase is defined to be $7= \lfloor 7 \rfloor_p$
and largest prime phase that is correctly identifiable is
$\lfloor 7^2 \rfloor_p$,
which is the next prime phase at the $N=3$rd
iteration. 
Hence the largest prime phase identifiable
at $N=3$ is $\lfloor \lfloor 7^2 \rfloor_p^2 \rfloor_p $,
which is the next prime phase at the step $N=4$.
The same argument applies at each iteration $N \geq 4$.
\end{proof}
\end{corollary}
The length of the
domain of phases over which all prime phases are correctly
inferred
therefore increases at each iteration as approximately
the square of the length of the previous domain,
and
is representable as a function of $7$
at each iteration. Recalling that $7$ is the largest
of the prime numbers derived at the first iteration
and that the derivation of the prime numbers is an initial
value problem, this reflects the dependence of the solution
on the initial conditions.

When applied to the results of Corollary $5.5$,
the Gauss/Legendre approximation 
$\pi(N)\approx N/ln(N)$
to the prime counting function, together with 
the approximation
$\lfloor N \rfloor_p \approx N$,  implies
\begin{corollary}
The maximum number of prime phases identifiable at each 
iteration of Theorem $ 5.3$ is approximately
${7^{2(N-1)}}/{(2(N-1)ln7)}$ for $N>1$.
\end{corollary}

There are computationally more-efficient
procedures than the recursive application of
$ {\overset {\bm \ast }{\bm  U}}_{N_{P}}(k)$ for identifying
prime phases. One has, for example,
\begin{corollary}
If $ p_1,  ...,, p_{N+1}$ are the first $N+1$ prime phases
then $p_N < k <p^2_{N+1}$ is
prime iff
$ 
\underset {j=1} {\overset{ N} \sum}
\overset {\bm \circ } 
{\bm {u}}(j) (k)=0.
$
\begin{proof}
By Theorem 5.3, the definition (\ref{eq:fdef}) of $\overset {\bm \circ } 
{\bm {u}}_{p_j}$, and the definition of the arithmetic plus operator. 
\end{proof}
\end{corollary}

\section    
{Representing Prime Phases as Functions
of the Initial Values}

While Theorem $5.3$ identifies the prime phases associated
with the circular product $	{\overset {\bm \ast }{\bm  U}}_{N_{P}}(k)$
of prime co-numbers, the representation of the phases
does not explicitly involve
the initial set of prime phases.
A more explicit representation may be obtained by employing
modular representations of the phases of the wave numbers.

\subsection{The Modular Representation Functions }

It is useful to define the modular phase function $\overset{m }{\bm u}_n$
of a wave number ${\bm u}_n$
as a function
that maps the frequency $k/n$ of non-zero terms 
to $(k\  mod\ n)/n$:
\begin{definition}
The principal sequence of values of the modular phase function 
$\overset{ m}{\bm u}_n$
of a wave number ${\bm u}_n$, represented as
frequencies of exponential terms, is
\begin{eqnarray} 
\label{eq:mpf}
 \ _1 \overset{ m}{\bm u}_n 
\sim
\left(  \frac{1\ mod\  n}{n}, \ \frac{2\  mod\  n}{n} ,..., 
\  \frac{n\  mod\  n}{n}   \right) 
=\left(  \frac{1}{n},  \ \frac{2}{n} ,...,
\  \frac{0 }{n}   \right) 
\end{eqnarray}
in which $\sim$ is to be interpreted as ``{\em is represented by}''.
\end{definition}
The principal sequence of function values
for the modular co-number is represented in terms of frequencies by
$ \ _1{\overset {\bm \ast m}{\bm u}_n} \sim 
\left(  1/{n}  ,..., \  (n-1)/{n}   ,\  \square  \right)$,
in which $\square$ represents that the corresponding function value is zero
rather than an exponential term.

The definition of the  circular product (\ref{eq:cdp}) must be modified 
for application to the values of modular phase functions
of natural wave numbers to
\begin{definition}
The modular circular product operation  $\overset{ m} {\bm u}_j 
\overset{ m} { \odot}   \overset{ m} {\bm u}_k  $
results in a sequence in which the frequency of the $k$th term is
\begin{eqnarray}
\label{eq:cdpa}
\frac{ \left(   P/n\  (k\  mod\  n) + P/m\  (k\  mod\  m)   \right)\ mod\  P}  {P},
\end{eqnarray}
in which $P=mn$
\end{definition}
For co-numbers $ {\overset {\bm \ast m}{\bm u}}_{2}$, 
$ {\overset {\bm \ast m}{\bm u}}_{3}$ in modular form, for example
\begin{eqnarray}
{_1}\left(\overset{\bm  \ast m}{\bm u}_2  \  
   \overset{ m} { \odot}\ \overset{\bm \ast m} {\bm u}_3 \right)\
&\sim& (1/2,\square,1/2,\square,1/2,\square)  \overset{ m} { \odot} (1/3,2/3,\square,1/3,2/3,\square) 
\nonumber\\
&=&((5\ mod\  6)/6,\square,\square,\square,(7\ mod\  6)/6,\square) \nonumber \\
 &=& ((5/6,\square,\square,\square,1/6,\square)  
\end{eqnarray}

\subsection{Modular Representations of the Prime Phases}

The modular wave co-numbers and modular circular product
may be used in defining the circular product
\begin{eqnarray}
\label{eq:CP01}
{\overset {\bm \ast m}{\bm U}}_{N_{P}}(k)=
{\underset {n=1} {\overset{ N} 
{ \overset m  \odot} }} {\overset {\bm \ast m}{\bm u}}_{p_n},
\hspace{.1in} for \  N=1,2,3,... 
\end{eqnarray}
which may be employed, rather than $ {\overset {\bm \ast } {\bm U}}_{N_P}(k)$,
in deriving representations of prime phases in terms of the
initial prime phases from which they are derived.
In the case for $N=3$ the function values for each $k$ are determined
by 
\begin{eqnarray}
\label{eq:correct}
{\overset {\bm \ast m}{\bm U}}_3(k)=
\frac{\left(p_2p_3(k\  mod\ p_1) + p_1p_3(k\  mod\  p_2)+p_1p_2(k\ mod\  p_3)\right)
mod\ p_1p_2p_3}
{p_1p_2p_3}
\end{eqnarray}
and are illustrated for $k=1,30$ in Table $I$,
from which it is clear that equation (\ref{eq:correct}) correctly represents
the value and phase of each of the prime numbers in this range
in terms of the prime numbers $(2,3,5)$ from which they are derived.
It is the case that they are correctly represented for $k<7^2$.

\begin{table}
  \centering
  \label{tab:branching}
  \begin{tabular}{|l|c|c|c|c|c|c|c|c|c|c|c|c|c|c|c|}
\hline                 \\
  ${\overset {\bm \ast m}{ \bm 2}} $   
&   $\tfrac{1}{2}$   &   $\square$   & $\tfrac{1}{2}$   
&  $\square$ &  $\tfrac{1}{2}$  &   $\square$   &   $\tfrac{1}{2}$   
&   $\square$  &   $\tfrac{1}{2}$   &  $\square$  &    $\tfrac{1}{2}$   &$\square$    
&  $\tfrac{1}{2}$  &  $\square$   &   $\tfrac{1}{2}$                   \\
\hline
  ${\overset {\bm \ast m }{ \bm3}} $      
&   $\tfrac{1}{3}$   &  $\tfrac{2}{3}$  &  $\square$   
&  $\tfrac{1}{3}$     &   $\tfrac{2}{3}$ &   $\square$   &$\tfrac{1}{3}$     
&  $\tfrac{2}{3}$   &   $\square$ &   $\tfrac{1}{3}$  &  $ \tfrac{2}{3}$   &   $\square$    
&  $\tfrac{1}{3}$   &   $\tfrac{2}{3}$  &  $\square$                  \\
\hline
$ {\overset {\bm \ast m}{ \bm 5}}   $   
&   $    \tfrac{1}{5}  $   &  $ \tfrac{2}{5}$   
&      $\tfrac{3}{5}$  &  $  \tfrac{4}{5}   $ &   $\square$   &   $\tfrac{1}{5} $   
&     $\tfrac{2}{5}$ &   $ \tfrac{3}{5} $  &   $ \tfrac{4}{5} $ & $\square$ 
&   $\tfrac{1}{5}$  & $ \tfrac{2}{5} \  $ &   $ \tfrac{3}{5} $    
&  $\tfrac{4}{5}$  &              $\square$                           \\
\hline
$ {\bm  {{\overset {\bm \ast m}{\bm U}}_N(k)}}\ $   
&   $\tfrac{1}{30}$   &   $\square$   
&  $\square$   &  $\square$  &  $\square$ &   $\square$   &   $\tfrac{7}{30}$   
&   $\square$ &   $\square$   &   $\square$  & $\tfrac{11}{30}$ &   $\square$    
&  $\tfrac{13}{30}$  &     $\square$    &    $\square$                   \\
\hline
\hline
$ {\overset {\bm \ast m }{ \bm 2}}  $  
&     $\square$     & $\tfrac{1}{2}  $   
&     $\square$ \ &  $  \tfrac{1}{2}   $ &   $\square$   &   $\tfrac{1}{2} $   
&    $\square$    &   $ \tfrac{1}{2}\  $ & $\square$  & $ \tfrac{1}{2} \  $ & $\square$    
&  $\tfrac{1}{2}$  &   $\square$   &     $\tfrac{1}{2}$   &   $\square$    \\
\hline 
$ {\overset {\bm \ast m }{ \bm 3}}   $  
&   $\tfrac{1}{3}$   &  $\tfrac{2}{3}$   & $\square$   
&  $\tfrac{1}{3}$  &  $\tfrac{2}{3}$    &   $\square$   &   $\tfrac{1}{3}$   
&   $\tfrac{2}{3}$  &   $\square$ &   $\tfrac{1}{3}$  & $\tfrac{2}{3}$ & $\square$    &  $\tfrac{1}{3}$  &       $\tfrac{2}{3}$   &    $\square$ \\
\hline   
$ {\overset {\bm \ast m }{ \bm 5}}   $   
&   $    \tfrac{1}{5}  $   &  $ \tfrac{2}{5}$   
&     $ \tfrac{3}{5}$  &  $  \tfrac{4}{5}   $ &   $\square$   &   $\tfrac{1}{5} $   
&    $ \tfrac{2}{5}$ &   $ \tfrac{3}{5} $  &   $ \tfrac{4}{5} $  &  $\square$  & 
$ \tfrac{1}{5} $  &  $\tfrac{2}{5}$   &   $\tfrac{3}{5}$    
&  $  \tfrac{4}{5}  $  &  $\square$ \\
\hline
$ {\bm  {{\overset {\bm \ast m}{\bm U}}_N(k)}}\ $       
&   $\square$   &   $\tfrac{17}{30}$  
&  $\square$  &   $\tfrac{19}{30}$      & $\square$   &   $\square$   
&   $\square$ &    $\tfrac{23}{30}$       &   $\square$  & $\square$   &    $\square$    
&  $\square$  &  $\square$   &      $\tfrac{29}{30}$   &  $\square$ \\
\hline
\hline
  \end{tabular}
$\vspace{.1in}$
\caption{Results for the computation of the circular products
of the first three modular prime co-numbers 
$\big(  \overset {\bm \ast m}{ \bm 2},\ 
  \overset {\bm \ast m }{ \bm 3},\ 
\overset {\bm \ast m }{ \bm 5}\big)$.
The symbol $\square$ represents a frequency corresponding to a zero occurring
in the principal phase of the prime wave co-numbers. The columns with only non-zero
frequencies represent prime wave numbers whose phases are derived
from the addition of the frequencies.}
\end{table}

Analogous computations of ${\overset {\bm \ast m}{\bm U}}_4(k)$
and ${\overset {\bm \ast m}{\bm U}}_5(k)$
lead to the identification of all prime phases less than
$p_{N+1}^2$ on the product circles, although they do not occur
in ascending order.
Hence one has the\\
{\bf Theorem 9}
{\em
If $p_1, ..., p_{N+1}$ are the first $N+1$ prime phases
and $P_N={\Pi_{j=1}^N p_j}$, 
then the set of phases $k=1,P_N $ that satisfy
the conditions
\begin{eqnarray}
\frac{k}{P_N} &=& \frac{1}{P_N} \left(
\left( P_N \left(\frac{k\  mod\ p_1}{p_1}+\frac{k\  mod\ p_2}{p_2}+...
+\frac{k\ mod\ p_N}{p_N}\right)\right)\ 
mod\  P_N \right),
\nonumber \\ 
{ k }\ &<\  &p^2_{N+1}
\end{eqnarray}
are all of the prime phases 
$ p_{N+1} \leq { k }\ <\  p^2_{N+1}$.
}

One notes possible connections between the expressions
representing the prime phases
and the zeta function $1/\zeta(1)$ for $s=1$.
For example,
since the total number of phases satisfying the first condition 
of Theorem $6.1$ is
${\Pi_{j=1}^N }(p_j-1)$,
the proportion of phases that must be considered 
as possible prime numbers is
\begin{eqnarray} 
\underset {n=1} {\overset{ N} \Pi} \left(\frac{(p_n-1)}{p_n}\right)
=1/\zeta_N(1)
\end{eqnarray}
in which $1/\zeta_N(1)$ represents the first $N$ terms
in the prime representation of the zeta function.

\section {Conclusions}

The natural wave numbers 
$
{\bm u}_n =\left(  e^{2\pi i ({k}/{n} ) } 
{ \tiny \ \ \vert \  k\ \epsilon\  \mathbb{Z}}, \ n\ \epsilon\  \mathbb{N}  \right)$
and their complements, the co-numbers 
$ 
{\overset{\bm  \ast} {\bm u}}_n =\left( 
(1-\delta_{k,jn})  e^{2\pi i ({k}/{n} )  } 
{ \tiny \ \ \vert\ j\ \epsilon\  \mathbb{Z}, \ 
k\ \epsilon\  \mathbb{Z}}, \ n\ \epsilon\  \mathbb{N}  \right)$,
provide
a useful representation of numbers that may be used for investigating
the distribution of the primes.
In particular, they
facilitate the identification of sets of numbers
that are defined in terms of not belonging to countable sets
of multiples of numbers that are specified in an initial set
of numbers.

The application of wave co-numbers to the identification
and representation of prime numbers leads to error-free predictions
of increasingly large sets of prime numbers that extend to all finite prime numbers.
These results, as presented in Theorem ${ 5.3}$ and its corollaries,
are conservative 
in predicting the next sequence of prime
numbers whose end-point
is determined by where the first error occurs for
an initial sequence of primes. 
Sequences of primes that follow
the square of the next known prime
have endpoints that may be
determined by more refined analyses.

The application also
leads, when the wave numbers are represented
in modular form, to representations
of prime numbers in terms of partial sums
of the reciprocals of the prime numbers
that form the data for the initial value problem
of identifying the primes.

Wavelike patterns in representations
of the primes that have been observed by many
investigators, and recently for example by
Wang (2021, 2022)\nocite{W21,W22},
follow naturally from a
representation of numbers in terms of sequences of the roots of unity.
It is reasonable to assume that the natural
wave numbers have applications in the physical sciences,
since a physical interpretation of the circular product
${{\bm u}}_m \odot {{\bm u}}_n= {{\bm u}}_{mn}$
is of three unit circles rotating at rates that are inversely proportional to
the number of their roots of unity.

\end{document}